# Fair Machine Learning for Healthcare Requires Recognizing the Intersectionality of Sociodemographic Factors, a Case Study

Alissa A. Valentine, MS [1], Alexander W. Charney, MD PhD [1], Isotta Landi, PhD [1]

[1] Icahn School of Medicine at Mount Sinai, New York, New York, USA

## Abstract

As interest in implementing artificial intelligence (AI) in medical systems grows, discussion continues on how to evaluate the fairness of these systems, or the disparities they may perpetuate. Socioeconomic status (SES) is commonly included in machine learning models to control for health inequities, with the underlying assumption that increased SES is associated with better health. In this work, we considered a large cohort of patients from the Mount Sinai Health System in New York City to investigate the effect of patient SES, race, and sex on schizophrenia (SCZ) diagnosis rates via a logistic regression model. Within an intersectional framework, patient SES, race, and sex were found to have significant interactions. Our findings showed that increased SES is associated with a higher probability of obtaining a SCZ diagnosis in Black Americans ($\beta = 4.1 \times 10^{-8}$, $SE = 4.5 \times 10^{-9}$, $p < 0.001$). Whereas high SES acts as a protective factor for SCZ diagnosis in White Americans ($\beta = -4.1 \times 10^{-8}$, $SE = 6.7 \times 10^{-9}$, $p < 0.001$). Further investigation is needed to reliably explain and quantify health disparities. Nevertheless, we advocate that building fair AI tools for the health care space requires recognizing the intersectionality of sociodemographic factors.

## 1 Introduction

Machine learning offers opportunities to develop AI tools for health care that reduce disparities and improve health outcomes for all. Yet, many models carry differential performance rates across patient sociodemographic groups, thus perpetuating disparity (Mehrabi et al., 2021; Huang et al., 2022; Segaret al., 2022; Kim and Cho, 2022). In response, interest continues to grow on the topic of generating fair, unbiased algorithms. This ushers in questions as to what variables should be considered when quantifying health disparities. Traditionally, socioeconomic status (SES) is used to account for inequities through quantifying education, occupation, income, and wealth (Braveman et al., 2010; Phelan et al., 2010; Berkowitz et al., 2015). In many instances, SES is included in models to explain differences in health outcomes between Black and White Americans (Williams and Jackson, 2005; Williams et al., 2010b, 2016b; Berkowitz et al., 2015; Nuru-Jeter et al., 2018). However, SES does not fully explain the racial inequities in American health care and the interaction between SES and race demands more attention (Williams et al., 2016a, 2010a; Keith and Brown, 2018; Bell et al., 2020).

In this paper, we focus on mental health research and explore the intersectional relationship between patient race, sex, and SES within the context of schizophrenia (SCZ). Intersectional perspectives recognize social categories like race and sex don't exist independently from each other and offer more complex investigations of disparity (Crenshaw, 1993). SCZ is a chronic and severe psychiatric illness that shows: (1) a racial disparity – the SCZ diagnosis rate is 2-4 greater in Black Americans than in White Americans (Strakowski et al., 2003; Blow et al., 2004; Bresnahan et al., 2007; Coleman et al., 2016; Olbert et al., 2018; Gara et al., 2019; Barr et al., 2022); (2) a sex disparity – the SCZ diagnosis rate is 1.4 times greater in males than in females (Aleman et al., 2003; Barr et al., 2022).

A longstanding generalization in psychiatry stipulates that higher SES at birth acts as a protective buffer against mental illnesses like SCZ, yet there is evidence this buffer is specific to White Americans in major depression disorder (Assari, 2017; Assari, 2017). There is also evidence that measures of SES mediate the racial disparity in SCZ diagnosis rates (Werner et al., 2007a,b; Fusar-Poli et al., 2016; Nagendra et al., 2020). However, the interplay of race, sex, and SES related to SCZ diagnosis in naturalistic settings has been less investigated. Our work proposes an intersectional framework to investigate the three-way interaction of patient race, sex, and SES in predicting SCZ diagnosis using electronic health record (EHR) data from the Mount Sinai Health System (MSHS), the largest medical system in New York City. To take into account known risk factors of SCZ, patient age and history of trauma or substance use disorder were included (Gut-Fayand et al., 2001; Aleman et al., 2003; Gearon et al., 2003; Seow et al., 2016; Setien-Suero et al., 2020).

Our findings demonstrate that race, sex, and SES influence a patient's risk of obtaining SCZ diagnosis, such that male Black Americans with high SES have the highest odds of obtaining a SCZ diagnosis ($OR = 1.022$), and male White Americans with high SES have the lowest odds of obtaining a SCZ diagnosis ($OR = 1.000$). This project contributes critical reflection on how to examine health disparities within a case study of SCZ. We suggest future research should consider:

- The context and epidemiology of a disorder, acknowledging power structures of institutional racism and bridging social justice with science;
- The complex interactions between sociodemographic variables that contribute to a patient's lived experienced with the medical system.

More importantly, we hope this work may act as a "call to action" for others in the field to acknowledge that variables like SES and race quantify disparity differently. Thus, creating models that omit one or the other may dismiss nuances which are important to recognize when aiming for model fairness.

## 2 Methods

In our study, we test the three-way interaction of patient race, sex, and SES in predicting SCZ diagnosis in consideration of known risk factors for SCZ such as age and history of trauma or substance use disorder.

### 2.1 Data

This dataset consists of real-world structured EHR data of all patients seen in the emergency department, inpatient, or outpatient settings between March 2006 and April 2023 of a large, urban health system. See Table 1 for an example of EHR data. The patients in these analyses are categorized into "the SCZ cohort" (i.e., patients with at least one SCZ diagnosis; $N = 12,105$) and "the control group" (i.e., patients with no SCZ diagnosis; $N = 2,506,838$). Data includes "patient-level values" (i.e., present throughout a patient's history) and "encounter-level values" (i.e., snapshots at specific time points from patient visits). Within both the SCZ and control cohort, patient-level values include date of birth (DOB), race and ethnicity, and sex. Both cohorts are limited to patients with (1) sex and reported as male or female and (2) race reported as Black or White. Other sex, gender, and racial or ethnic groups are not included due to insufficient sample sizes. See Table 2 for SCZ rates across included sociodemographic groups.

The SCZ cohort comprises of patients who received SCZ as a primary diagnosis between the ages of 13 and 98 years. A SCZ diagnosis was defined as any diagnosis that falls within the F20-29 International Classification of Diseases version 10 (ICD-10) codes. Age was then included in the model as the age at first SCZ diagnosis in the SCZ cohort. Within the SCZ cohort, the following encounter-level values were extracted from the patients' first encounter with a primary diagnosis of SCZ: encounter date/time, primary diagnosis, and zip-code at time of encounter. For the SCZ cohort, a history of trauma and substance use is attributed only to patients with a trauma/substance diagnosis before the SCZ diagnosis (see Table 1 for an example). Trauma and substance use diagnoses are defined as the ICD-10 codes mapped to corresponding Clinical Classifications Software Refined (CCSR) broader categories. SES is defined using patient zip code at first SCZ diagnosis, which is mapped to the United States Census Bureau for median-household income measures in 2021 that range from $2500-$250,000 (Census, 2021). The median household-income is a valid measure of patient SES and is related to health outcome disparities (Berkowitz et al., 2015).

Table 2: Cohort Description.

| Race | Sex | N | SCZ % |
|---|---|---|---|
| Black | Female | 397,713 | 0.8% |
| | Male | 303,383 | 1.4% |
| White | Female | 1,016,232 | 0.2% |
| | Male | 801,615 | 0.3% |

Table 1: Example EHR Data with Fictional Values.

| ID | Race | Sex | DOB | Visit Date | Primary Diagnosis | Zip-Code | Median Household Income |
|---|---|---|---|---|---|---|---|
| 1 | Black | Female | 1990-1-1 | 2021-3-3 | F20: Schizophrenia | 10029 | 33,901 |
| 2 | White | Male | 1980-1-1 | 2020-1-1 | F12: Cannabis Use | 10044 | 115,294 |
| 2 | White | Male | 1980-1-1 | 2020-2-2 | F20: Schizophrenia | 10044 | 115,294 |

The control cohort comprises of patients between 13 and 98 current years of age that have never received a SCZ diagnosis. Age was then included in the model as patients' current age in the control group. For the control cohort, a history of trauma and substance use attributed to patients with a trauma or substance use diagnosis. For calculating SES in control patients, zip-code is the one most often reported in encounter data.

## 2.2 Statistical Model

A logistic regression was conducted to investigate how patient race, sex, and SES interact and contribute to SCZ rates when controlling for known SCZ risk factors:

$$SCZ = Age + Susbtance\ Use\ Disorder + Trauma\ Disorder + Race \times Sex \times SES$$

The critical aspect of this model is the three-way interaction of patient race, sex, and SES which allows us to explore the different rates of SCZ diagnosis from an intersectional perspective.

## 3 Results

The regression was found to be statistically significant ($R^2 = 0.007$, $F(10, 2201973) = 1535$, $p < 0.001$). Significantly decreased rates of SCZ diagnosis were seen for older age [Mean age (SD) at first SCZ diagnosis = 51 years (20); $\beta = -4.8 \times 10^{-5}, SE = 2.5 \times 10^{-6}, p < 0.001$]. Significantly increased rates of SCZ diagnosis were seen for patients with a history of substance disorder ($\beta = 2.2 \times 10^{-2}$, $SE = 3.9 \times 10^{-4}$, $p < 0.001$; $OR = 1.022$) and trauma disorder ($\beta = 2.3 \times 10^{-2}$, $SE = 3.9 \times 10^{-4}$, $p < 0.001$; $OR = 1.023$). Black race ($\beta = 2.5 \times 10^{-3}$, $SE = 3.7 \times 10^{-4}$, $p < 0.001$) and male sex ($\beta = 2.8 \times 10^{-3}$, $SE = 3.4 \times 10^{-4}$, $p < 0.001$) were found to be associated with increased rates of SCZ. Higher SES was found to decrease rates of SCZ diagnosis ($\beta = -1.4 \times 10^{-8}$, $SE = 2.0 \times 10^{-9}$, $p < 0.001$).

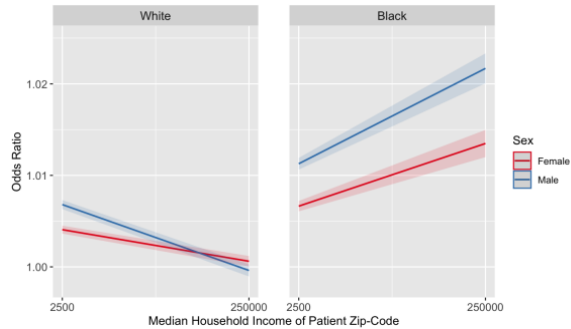

Figure 1: Odds Ratio of SCZ Diagnosis for varying median income by race and sex.

The significance of the two-way interaction terms showed that when controlling for other variables: Black male patients have increased risk of SCZ diagnosis ($\beta = 1.8 \times 10^{-3}$, $SE = 5.6 \times 10^{-4}$, $p < 0.01$); male patients with high SES have decreased risk of being diagnosed with SCZ ($\beta = -1.5 \times 10^{-3}$, $SE = 3.0 \times 10^{-9}$, $p < 0.001$); and Black patients with high SES have increased risk of being diagnosed with SCZ ($\beta = 4.1 \times 10^{-8}$, $SE = 4.5 \times 10^{-9}$, $p < 0.001$).

The three-way interaction term between race, sex, and SES was also significant ($\beta = 2.9 \times 10^{-8}$, $SE = 6.7 \times 10^{-9}$, $p < 0.001$). This result demonstrates that the interaction between race and sex differs across levels of SES such that male and female Black patients have increased odds of obtaining SCZ diagnosis if they are high SES compared to low SES when controlling for other risk factors (see Figure 1). As seen in Table 3, when comparing groups, Black male patients with high SES have the highest probability of being diagnosed with SCZ ($OR = 1.022$), and male patients with high SES have the lowest probability ($OR = 1.000$).

Table 3: Odds Ratio of SCZ Diagnosis by Race, Sex, and High or Low SES.

| Race | Sex | SES | SCZ % |
|---|---|---|---|
| Black | Female | High | 1.013 |
| | | Low | 1.007 |
| | Male | High | **1.022** |
| | | Low | 1.011 |
| White | Female | High | 1.001 |
| | | Low | 1.004 |
| | Male | High | **1.000** |
| | | Low | 1.007 |

## 4 Discussion

The findings from this project provides compelling evidence of how measures of SES interact with other sociodemographic features to contribute to health disparities. Within the context of SCZ, high SES serves as a protective buffer against diagnosis for White Americans, yet it appears to increase risk of diagnosis in Black Americans regardless of binary sex. This contradicts the generalized understanding that high SES is always beneficial to health outcomes. Our results also contribute to ongoing research arguing that variables like SES and race quantify disparity differently. This finding encourages critical thinking on what variables we prioritize when building machine learning models to mitigate health disparity. In the future, the best models should use utilize SES and race in an intersectional framework to evaluate health disparities.

## Ethics Statement

The team members of this project aim to acknowledge the limitations of this study as it prioritizes a sex/gender binary and does not recognize race or ethnicity outside of a Black-White binary. Additional work has been completed to explore SCZ diagnoses at the intersection of LGBTQ+ identity, race, and gender without SES being included, yet they are not included in this paper at this time but we hope to in the future. Furthermore, SES is a complex, intersectional variable on its own and our quantification of it is another limitation. In diverse, large urban areas specifically, there are large variations in income within the same zip code. Therefore, a more thorough assessment of patient SES would include both a neighborhood-level (e.g., median household income within zip code) and individual-level (e.g., insurance status) SES measure for each patient. It should be noted that our data is unique to our hospital system, and it's possible the distribution of patients diagnosed with SCZ across SES-racial groups varies in other health systems with specialized clinical programs. Nevertheless, these perspectives lead to questions as to how to quantify SES more granularly in diverse places.

## Acknowledgements

We thank Lili Chan, Ipek Ensari, and Ashwin Sawant for their invaluable contributions to this research as members of the first author's advisory committee. Special appreciation goes to the patients of Mount Sinai – may their data always be used for their benefit. We also acknowledge the funding and resources provided by the Mount Sinai Hospital System and Institute for Personalized Medicine.

## References

Andre Aleman, René S Kahn, and Jean-Paul Selten. Sex differences in the risk of schizophrenia: evidence from meta-analysis. Archives of general psychiatry, 60(6):565–571, 2003. ISSN 0003-990X.

Shervin Assari. High income protects whites but not african americans against risk of depression. In Healthcare, volume 6, page 37. MDPI. ISBN 2227-9032.

Shervin Assari. Social determinants of depression: The intersections of race, gender, and socioeconomic status. Brain sciences, 7(12):156, 2017.ISSN 2076-3425.

Peter B Barr, Tim B Bigdeli, and Jacquelyn L Meyers. Prevalence, comorbidity, and sociodemographic correlates of psychiatric disorders reported in the all of us research program. JAMA psychiatry, 2022.

C. N. Bell, T. K. Sacks, C. S. Thomas Tobin, and Jr. Thorpe, R. J. Racial non-equivalence of socioeconomic status and self-rated health among african americans and whites. SSM Popul Health, 10:100561, 2020. ISSN 2352-8273 (Print) 2352-8273. doi: 10.1016/j.ssmph.2020.100561. 2352-8273 Bell, Caryn N Sacks,Tina K Thomas Tobin, Courtney S Thorpe,Roland J Jr P2C HD041022/HD/NICHD NIH HHS/United States P2C HD041041/HD/NICHD NIH HHS/United States Journal Article England 2020/03/07 SSM Popul Health. 2020 Feb 21;10:100561. doi: 10.1016/j.ssmph.2020.100561.eCollection 2020 Apr.

Seth A Berkowitz, Carine Y Traore, Daniel E Singer, and Steven J Atlas. Evaluating area-based socioeconomic status indicators for monitoring disparities within health care systems: results from a primary care network. Health services research, 50(2):398–417, 2015. ISSN 0017-9124.

F. C. Blow, J. E. Zeber, J. F. McCarthy, M. Valenstein, L. Gillon, and C. R. Bingham. Ethnicity and diagnostic patterns in veterans with psychoses. Social Psychiatry and Psychiatric Epidemiology, 39 (10):841–851, 2004. ISSN 0933-7954. doi: 10.1007/s00127-004-0824-7.

Paula A Braveman, Catherine Cubbin, Susan Egerter, David R Williams, and Elsie Pamuk. Socioeconomic disparities in health in the united states: what the patterns tell us. American journal

Intersectionality of SES in Psychiatry

of public health, 100(S1):S186–S196, 2010. ISSN 1541-0048.

Michaeline Bresnahan, Melissa D Begg, Alan Brown, Catherine Schaefer, Nancy Sohler, Beverly Insel, Leah Vella, and Ezra Susser. Race and risk of schizophrenia in a us birth cohort: another example of health disparity? International Journal of Epidemiology, 36(4):751–758, 2007. ISSN 0300-5771. doi: 10.1093/ije/dym041. URL https://doi.org/10.1093/ije/dym041.

United States Bureau of the Census. 2021: American. community survey 1-year estimates subject tables. 2021.

Karen J Coleman, Christine Stewart, Beth E Waitzfelder, John E Zeber, Leo S Morales, Ameena T Ahmed, Brian K Ahmedani, Arne Beck, Laurel A Copeland, and Janet R Cummings. Racial-ethnic differences in psychiatric diagnoses and treatment across 11 health care systems in the mental health research network. Psychiatric Services, 67(7):749–757, 2016. ISSN 1075-2730.

K. Crenshaw. Mapping the margins - intersectionality, identity politics, and violence against women of color. Stanford Law Review Vol 43, No 6, July 1991, pages 1241–1299, 1993. URL <GotoISI>://WOS:A1993BZ75H00007. Bz75h

Paolo Fusar-Poli, Grazia Rutigliano, Daniel Stahl, André Schmidt, Valentina Ramella-Cravaro, Shetty Hitesh, and Philip McGuire. Deconstructing pretest risk enrichment to optimize prediction of psychosis in individuals at clinical high risk. JAMA Psychiatry, 73(12):1260–1267, 2016. ISSN 2168-622X. doi: 10.1001/jamapsychiatry.2016.2707. URL https://doi.org/10.1001/jamapsychiatry.2016.2707.

M. A. Gara, S. Minsky, S. M. Silverstein, T. Miskimen, and S. M. Strakowski. A naturalistic study of racial disparities in diagnoses at an outpatient behavioral health clinic. Psychiatric Services, 70 (2):130–134, 2019. ISSN 1075-2730. doi: 10.1176/appi.ps.201800223. URL <GotoISI>://WOS: 000462234300007. Hq2mb

J. S. Gearon, S. I. Kaltman, C. Brown, and A. S. Bellack. Traumatic life events and ptsd among women with substance use disorders and schizophrenia. Psychiatric Services, 54(4):523–528, 2003. ISSN 1075-2730. doi: DOI10.1176/appi.ps.54.4.523.

A. Gut-Fayand, A. Dervaux, J. P. Olie, H. Loo, M. F. Poirier, and M. O. Krebs. Substance abuse and suicidality in schizophrenia: a common risk factor linked to impulsivity. Psychiatry Research, 102(1): 65–72, 2001. ISSN 0165-1781. doi: Doi10.1016/S0165-1781(01)00250-5.

J. Huang, G. Galal, M. Etemadi, and M. Vaidyanathan. Evaluation and mitigation of racial bias in clinical machine learning models: Scoping review. JMIR Med Inform, 10(5): e36388, 2022. ISSN 2291-9694 (Print). doi: 10.2196/36388. 2291-9694 Huang, Jonathan Orcid: 0000-0002-3428-9952 Galal, Galal Orcid: 0000-0003-3525-859x Etemadi, Mozziyar Orcid: 0000-0002-6324-9220 Vaidyanathan, Mahesh Orcid: 0000-0003-4311-8896 Journal Article Review Canada 2022/06/01 JMIR Med Inform. 2022 May 31;10(5):e36388. doi: 10.2196/36388.

Verna M. Keith and Diane R. Brown. Mental Health: An Intersectional Approach, pages 131–142. Springer International Publishing, Cham, 2018. ISBN 978-3-319-76333-0. doi: 10.1007/ 978-3-319-76333-0 10. URL https://doi.org/10.1007/978-3-319-76333-0_10.

J. Y. Kim and S. B. Cho. An information theoretic approach to reducing algorithmic bias for machine learning. Neurocomputing, 500:26–38, 2022. ISSN 0925-2312. doi: 10.1016/j.neucom.2021.09.081.

N. Mehrabi, F. Morstatter, N. Saxena, K. Lerman, and A. Galstyan. A survey on bias and fairness in machine learning. Acm Computing Surveys, 54(6), 2021. ISSN 0360-0300. doi: Artn11510.1145/3457607.

Arundati Nagendra, Tate F. Halverson, Amy E. Pinkham, Philip D. Harvey, L. Fredrik Jarskog, Amy Weisman de Mamani, and David L. Penn. Neighborhood socioeconomic status and racial disparities in schizophrenia: An exploration of domains of functioning. Schizophrenia Research, 224:95–101, 2020. ISSN 0920-9964. doi: https://doi.org/10.1016/j.schres.2020.09.020. URL https://www.sciencedirect.com/science/article/pii/S0920996420304783.

Amani M Nuru-Jeter, Elizabeth K Michaels, Marilyn D Thomas, Alexis N Reeves, Roland J Thorpe Jr, and Thomas A LaVeist. Relative roles of race versus socioeconomic position in studies of health inequalities: a matter of interpretation. Annual review of public health, 39:169–188, 2018. ISSN 0163-7525.

C. M. Olbert, A. Nagendra, and B. Buck. Meta-analysis of black vs. white racial disparity in schizophrenia diagnosis in the united states: Do structured assessments attenuate racial disparities? J Abnorm Psychol, 127(1):104–115, 2018. ISSN 1939-1846 (Electronic) 0021-843X (Linking). doi: 10.1037/abn0000309. URL https://www.ncbi.nlm.nih.gov/pubmed/29094963. Olbert, Charles, M Nagendra, Arundati Buck, Benjamin eng Meta-Analysis 2017/11/03 J Abnorm Psychol. 2018 Jan;127(1):104-115. doi: 10.1037/abn0000309. Epub 2017 Nov 2.

Jo C Phelan, Bruce G Link, and Parisa Tehranifar. Social conditions as fundamental causes of health inequalities: theory, evidence, and policy implications. Journal of health and social behavior, 51 (1suppl) : S28 − −S40, 2010. ISSN 0022 − 1465.

Matthew W. Segar, Jennifer L. Hall, Pardeep S. Jhund, Tiffany M. Powell-Wiley, Alanna A. Morris, David Kao, Gregg C. Fonarow, Rosalba Hernandez, Nasrien E. Ibrahim, Christine Rutan, Ann Marie Navar, Laura M. Stevens, and Ambarish Pandey. Machine learning–based models incorporating social determinants of health vs traditional models for predicting in-hospital mortality in patients with heart failure. JAMA Cardiology, 7(8):844–854, 2022. ISSN 2380-6583. doi: 10.1001/jamacardio.2022.1900. URL https://doi.org/10.1001/jamacardio.2022.1900.

L. S. E. Seow, C. Ong, M. V. Mahesh, V. Sagayadevan, S. Shafie, S. A. Chong, and M. Subramaniam. A systematic review on comorbid post-traumatic stress disorder in schizophrenia. Schizophrenia Research, 176(2-3):441–451, 2016. ISSN 0920-9964. doi: 10.1016/j.schres.2016.05.004.

E. Setien-Suero, P. Suarez-Pinilla, A. Ferro, R. Tabares-Seisdedos, B. Crespo-Facorro, and R. Ayesa-Arriola. Childhood trauma and substance use underlying psychosis: a systematic review. European Journal of Psychotraumatology, 11(1), 2020. ISSN 2000-8198. doi: Artn174834210.1080/20008198.2020.1748342.

S. M. Strakowski, Jr. Keck, P. E., L. M. Arnold, J. Collins, R. M. Wilson, D. E. Fleck, K. B. Corey, J. Amicone, and V. R. Adebimpe. Ethnicity and diagnosis in patients with affective disorders. J Clin Psychiatry, 64(7):747–54, 2003. ISSN 0160-6689 (Print) 0160-6689 (Linking). doi: 10.4088/jcp.v64n0702. URL https://www.ncbi.nlm.nih.gov/pubmed/12934973. Strakowski, Stephen M Keck, Paul E Jr Arnold, Lesley M Collins, Jacqueline Wilson, Rodgers M Fleck, David E Corey, Kimberly B Amicone, Jennifer Adebimpe, Victor R eng MH56352/MH/NIMH NIH HHS/ Research Support, U.S. Gov't, P.H.S. 2003/08/26 J Clin Psychiatry. 2003 Jul;64(7):747-54. doi: 10.4088/jcp.v64n0702.

Shirli Werner, Dolores Malaspina, and Jonathan Rabinowitz. Socioeconomic status at birth is associated with risk of schizophrenia: Population-based multi-level study. Schizophrenia Bulletin, 33(6):1373–1378, 2007a. ISSN 0586-7614. doi: 10.1093/schbul/sbm032. URL https://doi.org/10.1093/schbul/sbm032. Shirli Werner, Dolores Malaspina, and Jonathan Rabinowitz. Socioeconomic status at birth is associated with risk of schizophrenia: population-based multi-level study. Schizophrenia bulletin, 33(6):1373–1378, 2007b. ISSN 1745-1701.

D. R. Williams, S. A. Mohammed, J. Leavell, and C. Collins. Race, socioeconomic status, and health: complexities, ongoing challenges, and research opportunities. Ann N Y Acad Sci, 1186: 69–101, 2010a. ISSN 0077-8923 (Print) 0077-8923. doi: 10.1111/j.1749-6632.2009.05339.x.1749-6632 Williams, David R Mohammed, Selina A Leavell, Jacinta Collins, Chiquita P01 AG020166/AG/NIA NIH HHS/United States U01 HL087322/HL/NHLBI NIH HHS/United States 3U01HL087322-02S1/HL/NHLBI NIH HHS/United States U-01 HL 87322-02/HL/NHLBI NIH HHS/United States Journal Article Research Support, N.I.H., Extramural Research Support, Non-U.S. Gov't United States 2010/03/06 Ann N Y Acad Sci. 2010 Feb;1186:69-101. doi: 10.1111/j.1749-6632.2009.05339.x.

D. R. Williams, N. Priest, and N. B. Anderson. Understanding associations among race, socioeconomic status, and health: Patterns and prospects. Health Psychol, 35(4):407–11, 2016a. ISSN 0278-6133 (Print) 0278-6133. doi: 10.1037/hea0000242. 1930-7810 Williams, David R Priest, Naomi Anderson, Norman B P50 CA148596/CA/NCI NIH HHS/United States P50 CA 148596/CA/NCI NIH HHS/United States Journal Article Research Support, N.I.H., Extramural United States 2016/03/29 Health Psychol. 2016 Apr;35(4):407-11. doi: 10.1037/hea0000242.

David R Williams and Pamela Braboy Jackson. Social sources of racial disparities in health. Health affairs, 24(2):325–334, 2005. ISSN 0278-2715.

David R Williams, Selina A Mohammed, Jacinta Leavell, and Chiquita Collins. Race, socioeconomic status, and health: complexities, ongoing challenges, and research opportunities. Annals of the New York Academy of Sciences, 1186(1):69–101, 2010b. ISSN 0077-8923.

David R Williams, Naomi Priest, and Norman B Anderson. Understanding associations among race, socioeconomic status, and health: Patterns and prospects. Health psychology, 35(4):407, 2016b. ISSN 1433823160